\documentclass[final,5p,times,twocolumn]{elsarticle}

\usepackage{amssymb}

\usepackage[slovak,english]{babel}
\usepackage[utf8]{inputenc}
\usepackage{booktabs}
\usepackage[separate-uncertainty=true, free-standing-units=true]{siunitx}
\DeclareSIUnit\eVc{\eV\per\clight\squared}
\DeclareSIUnit\clight{\text{\!\ensuremath{c}}}
\usepackage[version=4]{mhchem}
\newcommand{\cawo}{\ce{CaWO_4}}
\usepackage{multirow}
\usepackage{xcolor}
\usepackage{graphicx}
\usepackage{float}

\usepackage{cleveref}
\crefformat{appendix}{#2#1#3}

\journal{Applied Radiation and Isotopes}

\begin{document}


\begin{frontmatter}



\title{Secular Equilibrium Assessment in a \cawo{} Target Crystal from the Dark Matter Experiment CRESST using Bayesian Likelihood Normalisation}


\author[addrMPI]{G.~Angloher}
\author[addrHEPHY,addrAI]{S.~Banik}
\author[addrLNGS]{G.~Benato}
\author[addrMPI,addrCoimbra]{A.~Bento} 
\author[addrMPI]{A.~Bertolini} 
\author[addrBratislava]{R.~Breier}
\author[addrLNGS]{C.~Bucci} 
\author[addrHEPHY,addrAI]{J.~Burkhart\corref{cor1}}
\author[addrMPI]{L.~Canonica} 
\author[addrLNGS]{A.~D'Addabbo}
\author[addrLNGS]{S.~Di~Lorenzo}
\author[addrHEPHY,addrAI]{L.~Einfalt}
\author[addrTUM,addrWMI]{A.~Erb}
\author[addrTUM]{F.~v.~Feilitzsch} 
\author[addrMPI,addrClu]{N.~Ferreiro~Iachellini}
\author[addrHEPHY]{S.~Fichtinger}
\author[addrMPI]{D.~Fuchs} 
\author[addrHEPHY,addrAI]{A.~Fuss}
\author[addrMPI]{A.~Garai} 
\author[addrHEPHY]{V.M.~Ghete}
\author[addrLNGS]{P.~Gorla}
\author[addrHEPHY]{S.~Gupta} 
\author[addrMPI]{D.~Hauff} 
\author[addrBratislava]{M.~Je\v{s}kovsk\'y}
\author[addrTUE]{J.~Jochum}
\author[addrTUM]{M.~Kaznacheeva}
\author[addrTUM]{A.~Kinast}
\author[addrHEPHY]{H.~Kluck}
\author[addrOxford]{H.~Kraus} 
\author[addrTUM]{A.~Langenk\"amper} 
\author[addrMPI]{M.~Mancuso}
\author[addrLNGS,addrGSSI]{L.~Marini} 
\author[addrHEPHY]{V.~Mokina}
\author[addrMPI]{A.~Nilima} 
\author[addrLNGS]{M.~Olmi}
\author[addrTUM]{T.~Ortmann}
\author[addrLNGS,addrCASS]{C.~Pagliarone}
\author[addrLNGS,addrTUM]{L.~Pattavina}
\author[addrMPI]{F.~Petricca} 
\author[addrTUM]{W.~Potzel} 
\author[addrBratislava]{P.~Povinec}
\author[addrMPI]{F.~Pr\"obst}
\author[addrMPI]{F.~Pucci} 
\author[addrHEPHY,addrAI]{F.~Reindl} 
\author[addrTUM]{J.~Rothe} 
\author[addrMPI]{K.~Sch\"affner} 
\author[addrHEPHY,addrAI]{J.~Schieck} 
\author[addrHEPHY,addrAI]{D.~Schmiedmayer}
\author[addrTUM]{S.~Sch\"onert} 
\author[addrHEPHY,addrAI]{C.~Schwertner}
\author[addrMPI]{M.~Stahlberg} 
\author[addrMPI]{L.~Stodolsky} 
\author[addrTUE]{C.~Strandhagen}
\author[addrTUM]{R.~Strauss}
\author[addrTUE]{I.~Usherov}
\author[addrHEPHY]{F.~Wagner} 
\author[addrTUM]{M.~Willers} 
\author[addrMPI]{V.~Zema}
\author[addrLNGS,addrCop]{ \\ (CRESST Collaboration) \\ and \\ F.~Ferella}
\author[addrLNGS]{M.~Laubenstein}
\author[addrLNGS]{S.~Nisi}

\affiliation[addrMPI]{
organization={Max-Planck-Institut f\"ur Physik},
postcode={D-80805},
city={M\"unchen},
country={Germany}}

\affiliation[addrHEPHY]{
organization={Institut f\"ur Hochenergiephysik der \"Osterreichischen Akademie der Wissenschaften},
postcode={A-1050},
city={Wien},
country={Austria}}

\affiliation[addrAI]{
organization={Atominstitut, Technische Universit\"at Wien},
postcode={A-1020},
city={Wien},
country={Austria}}

\affiliation[addrLNGS]{
organization={INFN, Laboratori Nazionali del Gran Sasso},
postcode={I-67100},
city={Assergi},
country={Italy}}

\affiliation[addrBratislava]{
organization={Comenius University, Faculty of Mathematics, Physics and Informatics},
postcode={84248},
city={Bratislava},
country={Slovakia}}

\affiliation[addrTUM]{
organization={Physik-Department and ORIGINS Excellence Cluster, Technische Universit\"at M\"unchen},
postcode={D-85747},
city={Garching},
country={Germany}}

\affiliation[addrTUE]{
organization={Eberhard-Karls-Universit\"at T\"ubingen},
postcode={D-72076},
city={T\"ubingen},
country={Germany}}

\affiliation[addrOxford]{
organization={Department of Physics, University of Oxford},
postcode={OX1 3RH},
city={Oxford},
country={United Kingdom}}

\affiliation[addrCoimbra]{
organization={LIBPhys-UC, Departamento de Fisica, Universidade de Coimbra},
postcode={P3004 516},
city={Coimbra},
country={Portugal}}

\affiliation[addrWMI]{
organization={Walther-Mei\ss ner-Institut f\"ur Tieftemperaturforschung},
postcode={D-85748},
city={Garching},
country={Germany}}

\affiliation[addrClu]{
organization={Excellence Cluster Origins},
postcode={D-85748},
city={Garching},
country={Germany}}

\affiliation[addrGSSI]{
organization={GSSI-Gran Sasso Science Institute},
postcode={I-67100},
city={L'Aquila},
country={Italy}}

\affiliation[addrCASS]{
organization={Dipartimento di Ingegneria Civile e Meccanica, Universit\"a degli Studi di Cassino e del Lazio Meridionale},
postcode={I-03043},
city={Cassino},
country={Italy}}

\affiliation[addrCop]{
organization={Department of Physical and Chemical Sciences, University of l'Aquila, via Vetoio (COPPITO 1-2)},
postcode={I-67100},
city={L'Aquila},
country={Italy}}

\cortext[cor1]{Corresponding author: jens.burkhart@oeaw.ac.at}

\begin{abstract}
CRESST is a leading direct detection sub-\si{\giga\eVc} dark matter experiment. During its second phase, cryogenic bolometers were used to detect nuclear recoils off the \cawo{} target crystal nuclei. The previously established electromagnetic background model relies on Secular Equilibrium (SE) assumptions. In this work, a validation of SE is attempted by comparing two likelihood-based normalisation results using a recently developed spectral template normalisation method based on Bayesian likelihood. Albeit we find deviations from SE in some cases we conclude that these deviations are artefacts of the fit and that the assumptions of SE is physically meaningful.
\end{abstract}




\end{frontmatter}


\section{Introduction}
\label{sec:introduction}

The nature of Dark Matter (DM) is one of the most pre\-ssing problems of modern physics. Despite countless evidences from numerous astrophysical observations, no experiment could prove the presence of DM on earth thus far (see Ref.~\cite{Bertone:2004pz,ParticleDataGroup:2020ssz} for a review).

The direct detection experiment CRESST is one of the leading endeavours to probe the large parameter space of possible DM cross sections and particle masses in the sub-\si{\giga\eVc} range. The \cawo{}-based detector module TUM40, which this work focuses on and which was used during CRESST-II, had an energy detection threshold for nuclear recoils of \SI{603(2)}{eV}~\cite{angloher2014results}. This threshold was pushed down to \SI{30.1}{eV} with CRESST-III in a different detector module~\cite{CRESST:2019jnq}.

One of the limiting factors of the detection sensitivity of low mass DM experiments is the environmental radioactive background. In an earlier work, great efforts were made to simulate the most contributing sources of contamination in order to identify these backgrounds in the experimental data~\cite{abdelhameed2019geant4,abdelhameed2019err}. Recently, a normalisation method of the spectral templates based on Bayesian likelihood has been developed. The method shows great potential in improving the current background model. Details on this will be presented in a forthcoming paper. The goals of this work are to build upon this improved model and assess to what extend one can study secular equilibrium (SE) assumptions for the \cawo{} detector crystal.

The paper is structured as follows: \cref{sec:cressts-background-model} briefly summa\-rises the currently established background model of CRESST; \cref{sec:assessing-se} presents the methods used to assess SE and compares the fitting results; in \cref{sec:results-and-discussion}, partially broken SE is discussed; finally, \cref{sec:summary-and-outlook} concludes with a brief summary and an outlook.

\section{CRESST's Background Model}
\label{sec:cressts-background-model}

The considered contributions to CRESST's background can be separated into four categories:
\begin{itemize}
    \item Internal Radiogenic (IR) background. All 46 nuclides from the three natural radioactive decay chains (parent nuclides: $^{238}\mathrm{U}$, $^{235}\mathrm{U}$, and $^{232}\mathrm{Th}$, see \cref{fig:natural-activities}) and $^{40}\mathrm{K}$ inside the \cawo{} target crystal.
    \item Internal Cosmogenic (IC) background. Cosmogenic activation of the \cawo{} target crystal due to cosmic ray exposure during its production period. Thus far $^{179}\mathrm{Ta}$, $^{181}\mathrm{W}$, and $^{3}\mathrm{He}$ are considered.
    \item Near External Radiogenic (NER) background. All 46 nuclides from the three natural radioactive decay chains and $^{40}\mathrm{K}$ inside the copper holders surrounding the target crystal.
    \item Additional External Radiogenic (AER) background. Nine nuclides ($^{40}\mathrm{K}$, $^{208}\mathrm{Tl}$, $^{210}\mathrm{Pb}$, $^{212}\mathrm{Pb}$, $^{212}\mathrm{Bi}$, $^{214}\mathrm{Bi}$, $^{226}\mathrm{Ra}$ $^{228}\mathrm{Ac}$, and $^{234}\mathrm{Th}$) that can be clearly identified by their peaks in the experimental data, placed in a thin copper sphere enveloping the detector module (this is an approximation of the more distant copper components).
\end{itemize}
The nuclides are placed homogeneously in the corresponding source materials and their decays are simulated until a ground state is reached. The energy deposition distributions, which will be referred to as \textit{spectral templates} throughout the paper, are obtained for each parent nuclide.
In this work, we use the same templates as in Ref.~\cite{abdelhameed2019geant4,abdelhameed2019err}, where 72 million decays were simulated using the tool ImpCRESST, which is based on Geant4~\cite{agostinelli2003geant4} version 10.2 patch 1. In total, 85 templates\footnote{Templates with contributions completely outside of the fitting range or negligible expected contribution are disregarded.} are considered for the data fitting.

We consider an empirical detector response model to account for energy and time resolution of the detector:
\begin{itemize}
    \item Hits that happen within \SI{2}{ms} are counted as a single hit with an energy equal to the sum of the individual hits.
    \item Hit energies are randomly altered based on a Gauss distribution with energy-dependent spread.
\end{itemize}

We also use the same reference data as in Ref.~\cite{abdelhameed2019geant4,abdelhameed2019err} for the detector module TUM40.
The module's block-shaped \cawo{} crystal with a mass of \SI{246.2}{g} and a dimension of $32\times 32 \times 40$~\si{\mm\cubed} collected data for roughly \SI{528}{\day}, amounting to a total gross exposure of \SI{129.912}{\kg\day}.
The reference data is split into three data sets with labels based on the energy ranges they cover: \emph{low} [\SI{1}{keV}, \SI{495}{keV}], \emph{medium} [\SI{511}{keV}, \SI{2800}{keV}], and \emph{high} [\SI{4}{MeV}, \SI{7}{MeV}]. Additionally, the \emph{Region Of Interest} (ROI) marks the energy range [\SI{1}{keV}, \SI{40}{keV}], where the experiment has greater sensitivity of detecting DM\@.

\section{Discussion of Fit Results}
\label{sec:assessing-se}

In the previous work (see Ref.~\cite{abdelhameed2019geant4,abdelhameed2019err}), a data fitting method based on Secular Equilibrium (SE) assumptions\cite{l2012handbook} was used. In essence, the activities inside certain natural decay chains' sub-groups (SE groups) were assumed to be equal (apart from the branching ratios). However, the validity of these assumptions cannot be known a-priori due to the \cawo{} crystals' production spread. We therefore want to make a first assessment of the state of SE inside TUM40's target crystal since its background has been studied the most.

For this task, we perform two likelihood fits: one fit \emph{without} any SE assumptions (henceforth labelled \emph{uncorrelated fit}), i.e.\ 83 free-floating activity values, and one fit \emph{with} SE assumptions (henceforth labelled \emph{SE fit}), i.e.\ 52 free-floating activity values.

\begin{figure}[ht]
\centering
\includegraphics[width=1.0\linewidth]{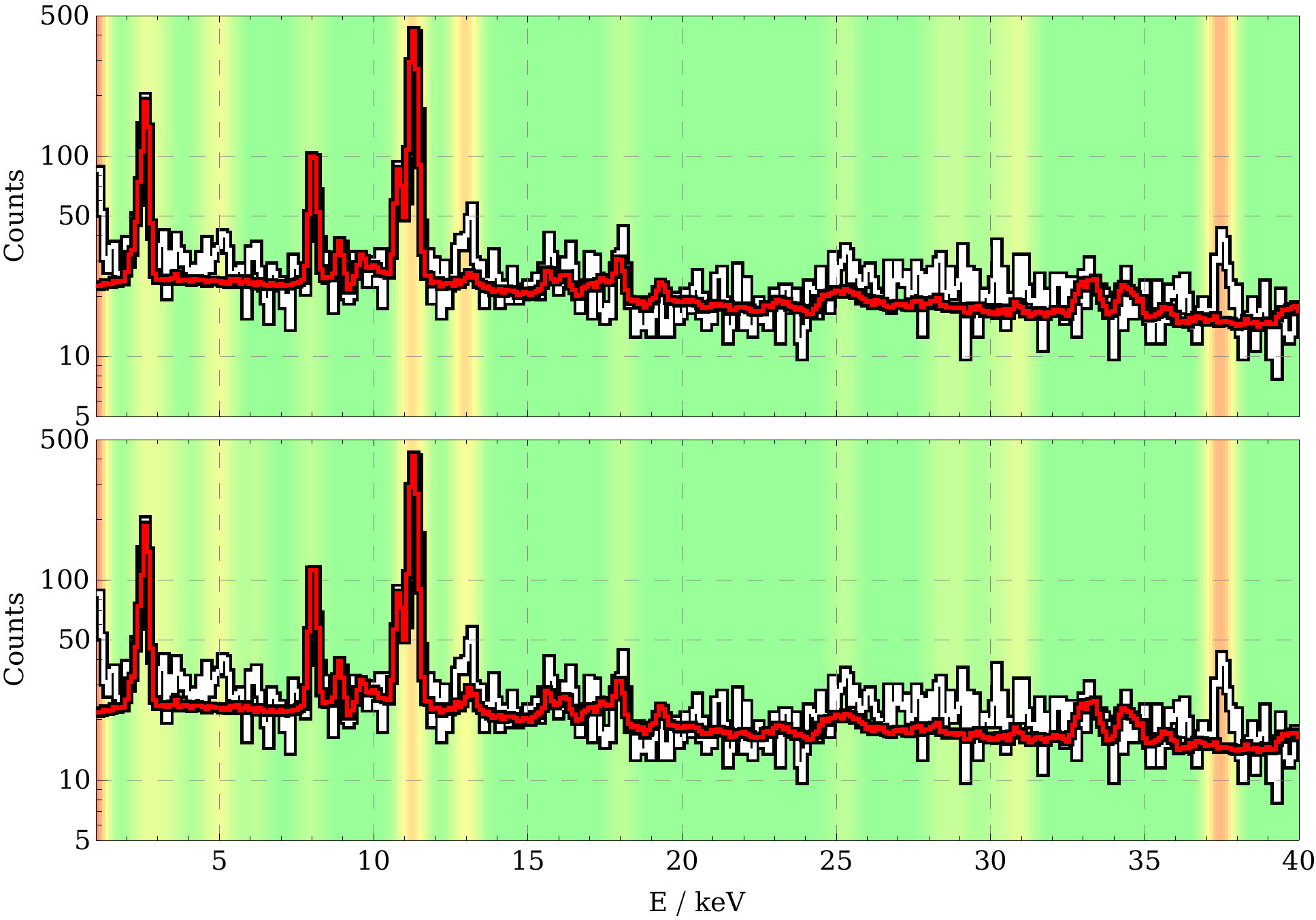}
\caption{Hypothesis plot comparing the performance of the uncorrelated fit (\textit{top}) and the SE fit (\textit{bottom}) in the ROI. The \textit{white line} represents the experimental data and the \textit{red line} represents the fit results. The \textit{background colour} shows the result of a hypothesis test with Weierstrass-smoothened~\cite{handbook2019zayed} colour values. Each bin is coloured according to the success (\textit{green}) or rejection (\textit{red}) of a null hypothesis test. The bin width is \SI{100}{eV}.}
\label{fig:hypothesis-roi}
\end{figure}

The results of the fits in the ROI can be seen in \cref{fig:hypothesis-roi}. 
Both results perform similarly well, despite the higher complexity of the uncorrelated fit. For comparison of fit results we use the \textit{coverage} of an energy range $\{ E \}$,
\begin{equation}
	\zeta_{\{ E \}} = \frac{N_{\text{MC}}}{N_\text{exp}} \biggr\rvert_{\{E\}} = \frac{\sum_{i=1}^{n_\text{bin}} \nu_j}{\sum_{i=1}^{n_\text{bin}} n_i},
	\label{eq:coverage}
\end{equation}
which is the ratio of the total fit counts $\nu$ over the experimental counts $n$, and the \textit{Explainable Percentage} ($\mathrm{EP}$)~\cite{burkhart2022enhancing},
\begin{equation}
	\text{EP}_{\{ E \}}=\frac{\sum_{i=1}^{n_\text{bin}} \Theta\left(p_c(n_i;\nu_i)-\alpha\right)\cdot n_i}{\sum_{i=1}^{n_\text{bin}} n_i},
	\label{eq:explainable-percentage}
\end{equation}
where $\Theta(x)$ is the Heaviside step function, and $p_c(n;\nu)$ is the central $p$-value~\cite{fay2010two, hirji2005exact},
\begin{equation}
	p_c(n;\nu)=\min(1, 2\cdot \min(p_l, p_r)),
	\label{eq:central-p-value}
\end{equation}
with the left- and right-sided $p$-values $p_l$ and $p_r$, respectively. In \cref{eq:explainable-percentage} the $\Theta(x)$ function acts as a hypothesis test with the nominal significance value $\alpha$, which was fixed to $\alpha=0.01$ before evaluation. The coverage measures the general over- or under-coverage of the fit while the $\mathrm{EP}$ takes the spectral distributions into account.

\begin{figure*}[ht]
\centering
\includegraphics[width=0.7\linewidth]{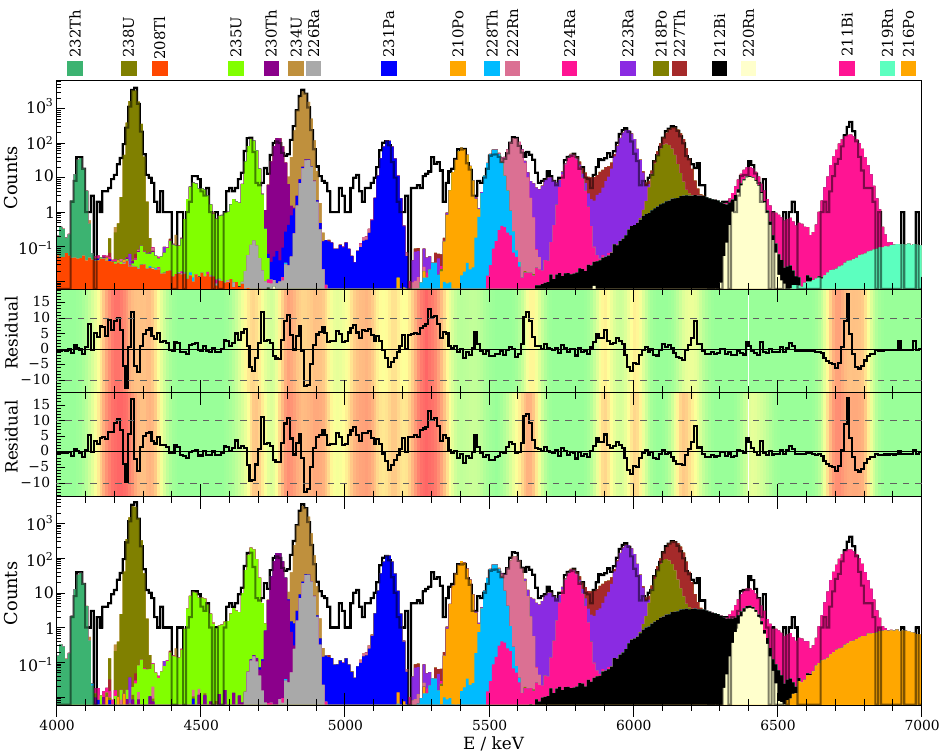}
\caption{Comparison of the uncorrelated fit (\textit{top}) and the SE fit (\textit{bottom}) in the high energy range. The \textit{back line} represents the data and the \textit{two middle plots} show the normalised residuals~\cite{gagunashvili2006comparison}, where the \textit{background colour} shows the result of a hypothesis test with Weierstrass-smoothened~\cite{handbook2019zayed} colour values. Each bin is coloured according to the success (\textit{green}) or rejection (\textit{red}) of a null hypothesis test. The bin width is \SI{10}{keV}.}
\label{fig:high-comparison}
\end{figure*}

Both the coverage $\zeta$~\cite{abdelhameed2019geant4,abdelhameed2019err} and the $\mathrm{EP}$ slightly favour the uncorrelated fit, except for the medium energy range, see \cref{tab:comparison-coverage-cenk-noSE-newLims}. 

\begin{table}[ht]
	\centering
	\begin{tabular}{l|cc|cc}
		\toprule
\textbf{Metric}     & \multicolumn{2}{c|}{$\zeta\, /\, \%$} & \multicolumn{2}{c}{$\mathrm{EP}\, /\, \%$} \\
\textbf{Fit}      & Uncorr. & SE & Uncorr. & SE \\
\midrule
\textbf{ROI}      & 90.3  & 89.4  & 81.7 & 81.4 \\
\textbf{Low}      & 93.4  & 93.7  & 87.3 & 86.9 \\
\textbf{Medium}   & 108.9 & 105.1 & 91.6 & 91.8 \\
\textbf{High}     & 100.1 & 97.3  & 41.5 & 32.3 \\
\textbf{Combined} & 98.6  & 97.2  & 75.9 & 73.3 \\
\bottomrule
	\end{tabular}
	\caption{Comparison of coverages ($\zeta$) and explainable percentages ($\mathrm{EP}$) between the uncorrelated fit (\textit{Uncorr.}) and the SE fit in the different energy ranges (see \cref{sec:cressts-background-model}).}
	\label{tab:comparison-coverage-cenk-noSE-newLims}
\end{table}
The highest tension in EP can be seen in the high energy range with a difference of \SI{9.2}{\percent} in favour of the uncorrelated fit. 
\Cref{fig:high-comparison} shows a comparison between the two fits in the high energy range.

Utilising the approximate Laplace integration method to obtain the evidence for each model, we get a Bayes factor~\cite{bayesFactors} of $>10^{10}$ in favour of the uncorrelated fit. 
However, it does not follow from this strong evidence against the SE fit that broken SE is prevalent. For example, an SE group's broad spectral templates are restricted by the group's peaky templates, while the unrestricted constituents can move independently (and unpenalised) such that broad, featureless spectra are able to compensate for contributions that are not yet implemented into the model.
A penalty term on deviation from SE and the inclusion of more radioactive sources might have a large influence on the Bayes factor. This will be studied in future works.

The deviation of both fits from the measured data can be either explained through an insufficient geometric setup implementation, missing background components like additional cosmogenically activated nuclides, or (partially) broken SE\@. While the first two options are currently under investigation, we focus now on laying foundations to study the latter option.

For this, we compare the activities of SE groups from the SE fit with the activities of their constituting nuclides from the uncorrelated fit. We consider here only the IR components since the NER components have comparatively low contributions and not enough AER components were simulated in Ref.~\cite{abdelhameed2019geant4,abdelhameed2019err} to infer their SE behaviour.

\section{Assessment of Broken Secular Equilibrium}
\label{sec:results-and-discussion}

\begin{figure*}[ht]
\centering
\includegraphics[width=1.0\textwidth]{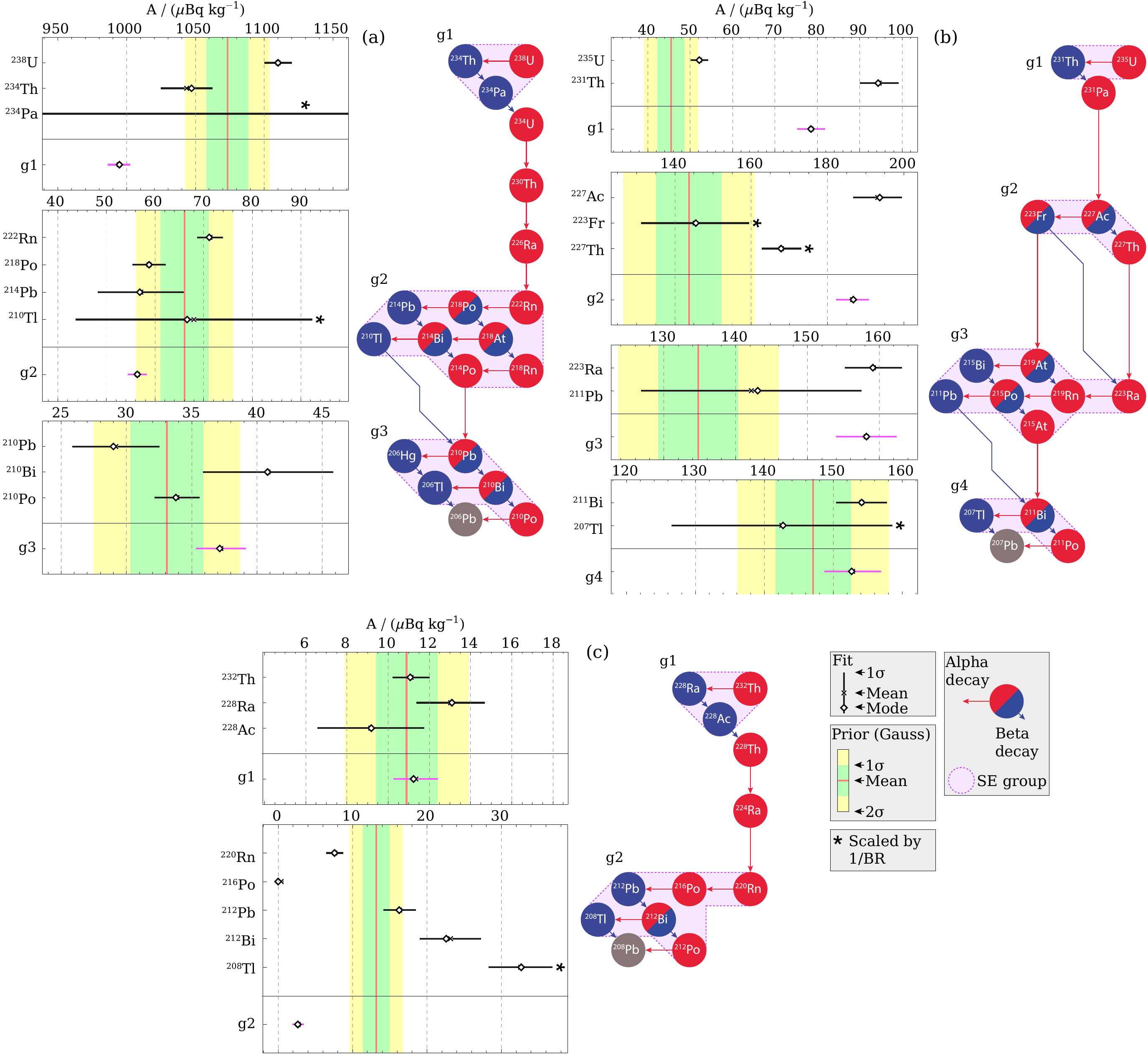}

\caption{SE assessment of IR components from (a) the $^{238}\mathrm{U}$ decay chain, (b) the $^{235}\mathrm{U}$ decay chain, and (c) the $^{232}\mathrm{Th}$ decay chain. The full decay chains are shown at the \textit{right sides}. 
At the \textit{left sides}, the SE groups' activities scaled to the respective parent nuclide contributions are shown as \textit{purple bars} (SE fit), while \textit{blue bars} show the activities of the groups' constituents (uncorrelated fit). Activities marked with an asterisk are scaled by the inverse branching ratio to make them comparable to their parent nuclides' activities. Similarly, the SE groups' activities are scaled to the groups' parent activities.}
\label{fig:natural-activities}
\end{figure*}

\Cref{fig:natural-activities} shows the activities of the IR components of all SE groups and compares them to the activities of the correlated SE groups. The priors for both fits were taken from the fit result in Ref.~\cite{abdelhameed2019geant4,abdelhameed2019err}\footnote{The influences of using these priors were studied and found to be negligible to the overall fit result while speeding up the fit convergence.} and are shown for comparison to the previous work.

The greatest discrepancies from SE occur in the first group of the $^{235}\mathrm{U}$ chain and in the second group of the $^{232}\mathrm{Th}$ chain. 
In the former case, the parent nuclide $^{235}\mathrm{U}$ has roughly half of the activity of its daughter $^{231}\mathrm{Th}$, while the SE group's activity lies in-between the two values. This is purely an effect of the fit, since $^{231}\mathrm{Th}$'s half-life of only $(25.52 \pm 0.01)\,\mathrm{h}$
makes a lasting contamination in the crystal production process unlikely.
In the latter case, the five activity values are scattered around the prior, while the SE group's activity lies well below the prior. This is, again, an effect of the fit because all nuclides of this group have a half-life around or less than \SI{1}{h}. Therefore, a contamination of daughter nuclides during the crystal's production process is unlikely.

The second SE group of the $^{235}\mathrm{U}$ chain exhibits slight discrepancy from SE due to effects of the fit. The daughters $^{223}\mathrm{Fr}$ and $^{227}\mathrm{Th}$ have lower activities than their parent $^{227}\mathrm{Ac}$, according to the uncorrelated fit, which is nonphysical.
The other two groups of the $^{235}\mathrm{U}$ chain are compatible with SE assumptions.

For the $^{232}\mathrm{Th}$ decay chain, the first group seems to be compatible with SE assumptions.

The observed discrepancies between the uncorrelated and the SE fit are to be expected for models of this magnitude. It is, on the other hand, noteworthy that most of the sub-groups \textit{are} compatible with SE assumptions -- even though the uncorrelated fit parameters could move through the parameter-space unrestricted from any physical penalty term. For the cases where strong deviations from SE are present, physically motivated reasons were given why SE must be upheld. We therefore conclude that SE assumptions hold and that the observed discrepancies are an artefact of the fit. The discrepancies must be either due to an insufficient geometric setup implementation or because background components are missing in the model.

\section{Summary and Outlook}
\label{sec:summary-and-outlook}

In this work we investigate the state of SE inside TUM40's \cawo{} target crystal using a likelihood-based normalisation method that scales the electromagnetic background components to the experimental reference data. Two fits are compared: one with rigid SE assumptions and one without any SE assumptions. While both fit results exhibit favourable metrics compared to the previous background model in Ref.~\cite{abdelhameed2019geant4,abdelhameed2019err}, the SE-constrained fit performs slightly poorer than the uncorrelated fit, which is to be expected. We then compare the activities of the two fit results to study what can be inferred about the upholding of SE inside the \cawo{} target crystal.

Overall, the SE assumptions can be validated by this investigation. Only two of the nine SE groups show deviation from SE\@. The reasons for these deviations are solely effects of the likelihood-normalisation. We argue that the half-lives of the deviating nuclides are too short to have a lasting effect on the data.

Both fit results show clear but similar deviations from the data, which can be attributed to a poor geometric implementation of the experimental setup and to a lack of considered contamination sources such as cosmogenically activated nuclides (see Ref.~\cite{kluck2021cosmic}). One example is the peak around \SI{37}{keV} that cannot be reconstructed in either model. We can therefore confidently say that the current set of simulated templates are not able to explain this peak -- even under the most loose parameter restrictions. This highlights why it is interesting to compare fits with SE assumptions to uncorrelated fits: Either a missing peak or feature in the background model is due to overly strict SE assumptions or the currently simulated spectral templates are ultimately not able to explain a certain peak, see \cref{fig:hypothesis-roi}.

After implementing a more detailed experimental geometry and considering more sources of radioactivity, we will also include data from the detector module Lise~\cite{angloher2014results} in future iterations of CRESST-II's background model. Especially the normalisation of AER, NER, and scintillating foil components will benefit from the additional data set. An assessment of SE -- like done in this work -- will follow with this more refined electromagnetic background model.

Furthermore, we will investigate a mixture between the two fitting modes presented here, namely to implement a penalty term in our likelihood function that constrains the fit towards SE while still allowing deviation, similar to what the DM experiment COSINE-100 uses~\cite{govinda2021cosine}.

Ultimately, the simulations presented here will help to obtain and justify a reliable background model that could then be used to subtract background signals in the current and new generations of the CRESST detectors. This would enhance the sensitivity of CRESST to potential DM signals.

\section{Acknowledgments}
This work has been funded by the Deutsche Forschungs\-gemeinschaft (DFG, German Research Foundation) under Germany's Excellence Strategy – EXC 2094 – 390783311 and through the Sonderforschungsbereich (Collaborative Research Center) SFB1258 ‘Neutrinos and Dark Matter in Astro- and Particle Physics’, by the BMBF 05A20WO1 and 05A20VTA and by the Austrian science fund (FWF): I5420-N, W1252-N27. FW was supported through the Austrian research promotion agency (FFG), project ML4CPD. SG was supported through the FWF project STRONG-DM (FG1). 
The Bratislava group acknowledges a partial support provided by the Slovak Research and Development Agency (project APVV-15-0576). The computational results presented were partially obtained using the CLIP cluster of the Vienna BioCenter and the Max Planck Computing and Data Facility (MPCDF).








\end{document}